\documentclass[12pt]{article}
\usepackage{epsfig}
\usepackage{amsmath,amssymb,amsthm,amscd}
\def\be{\begin{equation}}
\def\ee{\end{equation}}
\newcommand{\sectiono}[1]{\section{#1}\setcounter{equation}{0}}

\def\ha{\hat a}
\def\hb{\hat b}

                                                     \newcommand{\figref}[1]{Fig.~\protect\ref{#1}}
\begin{document}
{}~
\hfill\vbox{
\hbox{ITFA-2005-39}
\hbox{CERN-PH-TH/2005-156}
\hbox{NSF-KITP-05-69}
}\break

\vskip 1cm
\centerline{\Large \bf
Phase transitions in $q$-deformed 2D Yang-Mills theory}\vskip .1cm
\centerline{\Large \bf and topological strings}

\vspace*{5.0ex}

\centerline{ \large \rm Xerxes Arsiwalla$^a$, Rutger Boels$^a$,
Marcos Mari\~no$^b$\footnote{Also at Departamento de Matem\'atica, IST, Lisboa, Portugal} and Annamaria Sinkovics$^a$}

\vspace*{4.0ex}

\centerline{ \rm ~$^a$Institute for Theoretical Physics, University of Amsterdam  }

\centerline{ \rm Valckenierstraat 65, 1018 XE Amsterdam, The Netherlands}

\vspace*{1.8ex}

\centerline{ \rm ~$^b$Department of Physics, Theory Division, CERN}

\centerline{ \rm CH-1211 Geneva, Switzerland}

\vspace*{6ex}

\centerline{\bf Abstract}
\medskip

We analyze large $N$ phase transitions for $U(N)$ $q$-deformed two-dimensional Yang-Mills theory on the sphere. We determine the
phase diagram of the model and we show that, for small values of the deformation parameter, the theory exhibits a phase transition which is smoothly
connected to the Douglas-Kazakov phase transition. For large values of the deformation parameter the phase transition
is absent. By explicitly computing the one-instanton suppression factor in the weakly coupled phase,
we also show that the
transition is triggered by instanton effects. Finally, we present the solution of the model in the strongly coupled phase. Our analysis
suggests that, on certain backgrounds, nonperturbative topological string theory has new phase transitions at small radius. From the point
of view of gauge theory, it suggests a mechanism to smooth out large $N$ phase transitions.

\vfill \eject

\baselineskip=16pt

\tableofcontents
\sectiono{Introduction}

In certain cases, nonperturbative completions of string theory
can be obtained by considering a holographic description
in terms of a D-brane gauge theory. Recently, in the case of topological strings,
Ooguri, Strominger and Vafa \cite{OSV} made a proposal for such a nonperturbative
completion based on the connection
 to the black hole attractor mechanism. According to \cite{OSV}, the nonperturbative description of topological string theory on a
 Calabi-Yau background is encoded in a D-brane gauge theory living on some appropriate
 cycles of the manifold.

 In \cite{V,AOSV} this proposal was made more concrete by considering Calabi-Yau backgrounds
 of the form
 \be
 L_1 \oplus L_2 \rightarrow \Sigma_g,
 \ee
 where $\Sigma_g$ is a Riemann surface of genus $g$ and $L_1, L_2$ are line bundles such that ${\rm deg}(L_1) + {\rm deg}(L_2)=2g-2$. In this case,
 the relevant D-brane gauge theory reduces
 to a $q$-deformed version of two-dimensional Yang-Mills (YM) theory on the Riemann surface $\Sigma_g$. $q$-deformed 2d YM can be regarded as a one-parameter deformation of the standard 2d YM theory. As we will explain below, the deformation can be parametrized by a real, positive number $p$, in such a way that as $p\rightarrow \infty$ one recovers the standard YM theory. The $q$-deformed theory is exactly solvable and one can compute its partition function on any Riemann surface. This partition function has a strong coupling expansion as a sum over representations of the gauge group, which can be written, following \cite{GT}, in terms of a product of a chiral and an antichiral sector. The perturbative topological string partition function, which was computed in \cite{BP} for this class of geometries, is given by a certain limit of this expansion in which the antichiral sector decouples. Once we have a nonperturbative description of the theory, it is natural to ask what new phenomena emerge in this description and what their implications are for string theory. For example, in \cite{DGOV} the fermionic description of 2d YM on the torus was used to study baby universes in string theory.

2d YM theory on the sphere exhibits an interesting phenomenon: as shown by Douglas and Kazakov \cite{DK}, there is a large $N$, third order phase transition at a critical value of the area $A=\pi^2$ between a large area phase and a small area phase. From the point of view of the small area/weak coupling phase, the phase transition is triggered by instantons \cite{GM}. From the point of view of the large area/strong coupling phase and its string description in terms of branched coverings \cite{G, GT}, the transition is triggered by the entropy of branch-point singularities \cite{T}.
Due to the Douglas-Kazakov transition, the large area expansion of 2d YM theory on the sphere has a finite radius of convergence \cite{T}.

In this paper we will study the possibility of large $N$ phase transitions in $q$-deformed 2d YM. Since as the deformation
parameter $p$ goes to infinity we recover the usual theory, it is
natural to expect the transition to occur at large enough values of $p$. In fact, our result show that the
transition persists for all $p>2$, and we find a critical line smoothly connected to the Douglas-Kazakov transition of the standard 2d YM theory. We also show that for $p\le 2$, in the regime of strong $q$-deformation, the phase transition does not occur. We also perform a detailed instanton analysis which shows that, as in the standard YM case studied in \cite{GM},  the transition is triggered by instanton effects.

Most of the analysis of this paper are done in the small area phase. In 2d YM theory this phase is described by a Gaussian matrix model. In the $q$-deformed case, this phase is essentially described by the Chern-Simons or Stieltjes-Wigert matrix model introduced in \cite{MM} and
studied in \cite{AKMV,T,MMr}\footnote{Connections between Chern-Simons theory and $q$-deformed 2d Yang-Mills theory have been made, from a
different perspective, in \cite{AOSV} and \cite{HT}.}. This model, albeit complicated, is exactly solvable (in terms of, for example, orthogonal polynomials), and this is the underlying reason that we can make exact statements
about the location of the critical line and the instanton contributions. The large area phase turns out to be more difficult to handle. In this paper we present some preliminary results and derive the equations that determine the full solution (including an explicit expression for the two-cut
resolvent). We expect the phase transition of the $q$-deformed theory to be of third order for $p>2$, since it is smoothly connected to the transition of
Douglas and Kazakov, and indeed we give indirect evidence that this is so.

As in the standard 2d YM, the existence of the phase transition in the $q$-deformed version indicates that the large area expansion has
a finite radius of convergence. According to \cite{T,AOSV}, this theory provides a nonperturbative description of topological string theory on certain Calabi-Yau backgrounds. This suggests that the large area expansion breaks down in the full topological string theory, and there is a phase transition between a small area phase and a large area phase. From the gauge theory point of view, our analysis shows that when the $q$-deformation is strong enough, the model exhibits a single phase. This suggests that $q$-deformations give a mechanism to smooth out large $N$ phase transitions.

The structure of this paper is as follows: in section 2 we briefly review the Douglas-Kazakov transition in 2d YM theory. In section 3 we determine the
phase diagram of the $q$-deformed theory and we find a line of critical points parametrized by $p$, for $p>2$. In section 4 we adapt the
analysis of \cite{GM} and study the
phase transition of the $q$-deformed theory in terms of instantons in the weakly coupled phase. We find an explicit expression for the one-instanton suppression factor which indicates that, indeed, the transition is triggered by instanton effects. In section 5 we analyze the large area phase, which can be encoded by standard techniques in a two-cut solution to an auxiliary matrix model. Finally, in section 6, we discuss the implications of our results for topological string theory and outline some problems opened by this investigation.

As this paper was being completed, we became aware of the work  \cite{MY}, where the same problem is studied. After submission, another paper appeared \cite{Italians} addressing the 
same issues. 

\sectiono{A review of the Douglas-Kazakov transition}

2d YM theory is an exactly solvable model. In particular, the partition function
of the $U(N)$ theory on the sphere is given by a sum over representations of $U(N)$ (see \cite{cmr} and
references therein)
\be
\label{stpf}
Z=\sum_R \bigl( {\rm dim}\, R\bigr)^2 e^{-A C_2(R)/2N} e^{i \theta C_1(R)},
\ee
where ${\rm dim}\, R$ is the dimension of the representation $R$, $A$ is a real and positive parameter that can be identified with the area of the
sphere, and $C_1(R)$, $C_2(R)$ are the first and second Casimir of $R$. We will represent $R$ by a
set of integers $\{ l_1, l_2, \cdots, l_N\}$ satisfying the inequality
\be
\label{ineq}
\infty \ge l_1 \ge l_2 \ge \cdots \ge l_N \ge -\infty.
\ee
In terms of these integers, the Casimirs have the expression
\be
\begin{aligned}
C_1(R)=&\sum_{i=1}^N l_i,\\
C_2(R)=&\sum_{i=1}^N l_i(l_i -2 i + N+1).
\end{aligned}
\ee
Although the above partition function looks rather simple, this theory turns out to have a very
rich structure. In \cite{G,GT} it was shown that at large area the partition function (\ref{stpf})
admits a string representation
in terms of branched coverings of Riemann surfaces (see \cite{cmr} for an excellent review).
Douglas and Kazakov found that the planar free energy on the sphere
exhibits a third order phase transition at the critical value
\be
\label{acrit}
A_*=\pi^2.
\ee
This large $N$ transition is a continuum analogue of the Gross-Witten-Wadia phase transition
for 2d YM theory on the lattice \cite{GW,W}. Since in this paper we will be considering a generalization of the Douglas-Kazakov phase transition, we will briefly review how this transition is found. For the rest of this section we will set $\theta=0$.

At large $N$ it is natural to introduce a distribution of Young tableaux
\be
n(x)={l_i\over N}, \quad x={i\over N}.
\ee
Defining the shifted distribution
\be
\label{hx}
h(x)=-n(x) + x-{1\over 2},
\ee
one finds that the planar free energy is given
by
\be
\label{sf}
F_0(A) =-S_G[h],
\ee
where the functional $S_G[h]$ reads
\be
\label{shg}
S_G[h]=-\int_0^1 dx \int_0^1 dy \log | h(x)-h(y)| + {A \over 2} \int_0^1 dx h(x)^2 -{A \over 24} -{3\over 2}.
\ee
Let us now introduce the density function
\be
\label{hdens}
\rho(h)={dx\over dh},
\ee
which is normalized to unity,
\be
\int dh \, \rho(h)=1.
\label{densnorm}
\ee
One crucial observation of \cite{DK} is that, because of the inequality (\ref{ineq}), this density has to
satisfy
\be
\rho(h)\le 1
\label{rineq}
\ee
for all $h$. We can now write (\ref{shg}) as
\be
\label{actiong}
S_G[\rho]= -\int dh \int dh' \, \rho(h) \rho(h') \log | h-h'| + {A\over 2} \int dh \rho(h)h^2 -{A \over 24}
-{3\over 2}.
\ee
This is (up to the $\rho$-independent terms) the saddle-point functional for a Gaussian matrix model
with 't Hooft parameter $t=1/A$. It then follows that the density $\rho(h)$ is given by Wigner's semicircle law,
\be
\rho_G(\lambda, t)={1\over 2 \pi t} {\sqrt { 4t - \lambda^2}},
\ee
and we find
\be
\rho(h)=\rho_G(h, 1/A).
\ee
However, it is clear that this solution can be valid only for $A\le \pi^2$, since after this point
the inequality (\ref{rineq}) is violated. This indicates that there is a phase transition at the critical value
(\ref{acrit}).
\begin{figure}[!ht]
\leavevmode
\begin{center}
\epsfysize=4cm
\epsfbox{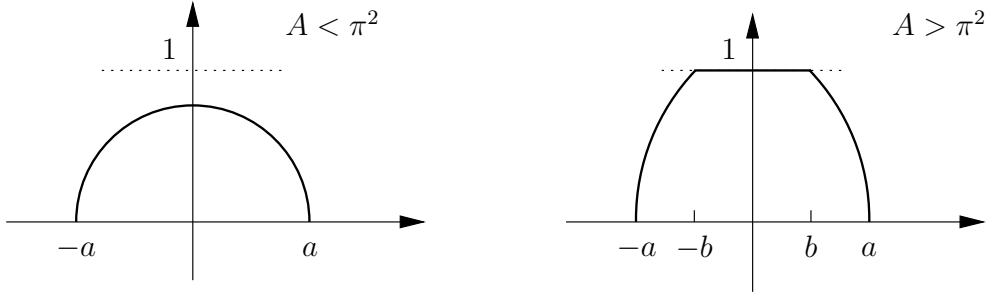}
\end{center}
\caption{This figure shows the density $\rho(h)$ before and after
the Douglas-Kazakov transition. The solution for $A\ge \pi^2$ can be interpreted as a two-cut solution
of an auxiliary matrix model.}
\label{dkt}
\end{figure}

For $A\ge \pi^2$ the Gaussian solution is no longer valid, and Douglas and Kazakov argued
that one could obtain a solution for the large area phase by considering a a density of eigenvalues
of the form,
\be
\label{tcrho}
\rho(h)=\begin{cases} \tilde \rho(h), &\text{$-a \le h \le -b$,  $b \le h\le a$}, \\
1, &\text{$-b \le h \le b$,}
\end{cases}
\ee
where $b<a$ are points in the real positive axis. From the point of view of the density $\rho(h)$, the Douglas-Kazakov transition can be represented as in \figref{dkt}: for $A<\pi^2$ the Gaussian density
gives a good description, but as $A\ge \pi^2$ one finds a new density of the form (\ref{tcrho}). It is easy to see that finding $\tilde \rho(h)$ amounts to finding a two-cut solution for
a modified matrix model with a logarithmic potential. The explicit solution to this problem was worked out in \cite{DK}, and this allowed them to verify that the phase transition at $A=\pi^2$ is of third order. It
was also verified that the large area solution agrees with the string expansion of \cite{GT}.

The mechanism behind the Douglas-Kazakov phase transition was further elucidated in \cite{MP,GM, CD}.
In particular, it was shown by Gross and Matytsin in \cite{GM} that the Douglas-Kazakov phase transition is driven by instantons. The small area phase is dominated by the perturbative vacuum, and instantons are suppressed with an $\exp(-N)$ factor. The one-instanton suppression factor at leading
order in $N$ was computed in
\cite{GM} to be given by
\be
\exp\biggl[ -{N \over A} \gamma_{\rm GM}(A)\biggr],
\ee
where
\be
\label{grossm}
\gamma_{\rm GM}(A)=2\pi  {\sqrt {\pi^2 -A}} -A \log\biggl[ {(\pi + {\sqrt {\pi^2 -A}})^2 \over A}\biggr] .
\ee
Since $\gamma_{\rm GM}(A=\pi^2)=0$, as we reach the critical point instantons are not anymore suppressed and
they trigger the phase transition, which is then a consequence of $\exp(-N)$ effects which are not
visible in the $1/N$ expansion.

\sectiono{The phase diagram of $q$-deformed 2d YM}

The $q$-deformed two-dimensional Yang-Mills theory arises as a natural deformation
of the usual model. This model has been considered in \cite{BR, KL} and more recently, in the context of topological string theory, in \cite{AOSV}. The partition function of the $q$-deformed theory on the sphere can be obtained by replacing the dimensions of representations in (\ref{stpf}) by
their quantum counterpart, in the sense of quantum group theory. The resulting partition function
depends on the rank $N$ of the gauge group, two real parameters, $p, g_s$, and an angle $\theta$. It reads,
\be
Z^{q}=\sum_R \bigl( {\rm dim}_q R\bigr)^2 q^{p C_2(R)/ 2} e^{i \theta C_1(R)},
\label{zq}
\ee
where the quantum dimension of $R$ is given by
\be
{\rm dim}_q R=\prod_{1\le i<j\le N} {[l_i -l_j+j-i]\over [j-i]},
\ee
and the $q$-numbers appearing here are defined as
\be
[x]=q^{x\over2} -q^{-{x\over2}}, \quad q=e^{-g_s}.
\ee
The free energy of the model is defined as
\be
F^q={1\over N^2} \log \, Z^q.
\ee
It is convenient to define the parameter $A$ as
\be
p\, g_s ={A\over N}.
\ee
As we will see in a moment, $A$ corresponds to the area of the sphere in (\ref{stpf}). As in 2d YM, we will require $A$ to be positive. Notice that the $q$-deformed theory is symmetric under $p, g_s \rightarrow -p, -g_s$. Therefore,
we can restrict ourselves to the range of parameters $p>0$, $g_s>0$.

An important property of the $q$-deformed theory is that in a suitable double-scaling limit, one recovers ordinary 2D YM. This limit is defined
as follows:
\be
p \longrightarrow \infty, \qquad g_s \longrightarrow 0, \qquad A, \, N \, \, \, {\rm fixed}.
\label{dslone}
\ee
As $g_s \rightarrow 0$ with
$N$ fixed, the quantum dimension becomes the classical dimension:
\be
{\rm dim}_q R \longrightarrow {\rm dim}\, R,
\ee
and
\be
q^{p C_2(R) /2} \longrightarrow \exp\biggl( -{A C_2(R)\over 2 N} \biggr),
\ee
which is the standard weight factor for 2d YM. We then recover the partition function (\ref{stpf}) for a sphere of area $A$. The $q$-deformed theory can then
be regarded as a one-parameter deformation of 2d YM.

In this paper we will  be interested in the large $N$ dynamics of the deformed theory. It is useful to
introduce the 't Hooft parameter, which is defined as
\be
t=N g_s,
\ee
and we will consider the 't Hooft large $N$ limit in which $N \rightarrow \infty$ and $t$ and $p$ are fixed. The planar
free energy
\be
F^q_0(t,p) =\lim_{N \to \infty} F^q
\ee
 will then be a function of $t$ and $p$. Notice that the limit (\ref{dslone}) that
gives ordinary Yang-Mills theory can be implemented order by order in the $1/N$ expansion
by taking
\be
\label{dsl}
p\longrightarrow \infty,  \qquad t\longrightarrow 0, \qquad pt =A \, \, {\rm fixed}.
\ee
In this way, we recover planar 2d YM on the sphere. We will check many of our results for the $q$-deformed theory, by
verifying that in the limit (\ref{dsl}) one recovers the known results in 2d YM.

In order to compute the planar free energy, we follow the steps outlined in the previous section for the undeformed theory and
represent the planar free energy in terms of a functional of a distribution $h(x)$, which is
defined as in (\ref{hx}). It is easy to see that in the large $N$ limit the planar free energy derived from (\ref{zq}) is given by
\be
\label{pf}
F^q_0(t,p) =-S[h],
\ee
where the functional $S[h]$ reads
\be
\begin{aligned}
\label{spf}
S[h]=&-\int_0^1 dx \int_0^1 dy \log | 2 \sinh \, {t\over 2} (h(x)-h(y))| + {p t \over 2} \int_0^1 dx h(x)^2 \\
& + i \theta \int_0^1 dx h(x) -{pt \over 24} + \int_0^1 dx \int_0^1 dy \log | 2 \sinh \, {t\over 2} (x-y)|,
\end{aligned}
\ee
and in (\ref{pf}) $S[h]$ is evaluated on the configuration $h(x)$ which minimizes the above functional. The last term in (\ref{spf}) comes from the denominator of the quantum dimension and it is given by
\be
\int_0^1 dx \int_0^1 dy \log | 2 \sinh \, {t\over 2} (x-y)|={2\over t^2} F_0^{\rm CS}(t),
\ee
where
\be
F^{\rm CS}_0(t)= {t^3 \over 12} - {\pi^2 t \over 6} -{\rm Li}_3(e^{-t}) + \zeta(3).
\ee
This function is the planar free energy of Chern-Simons theory \cite{GV}, and we recall that
the polylogarithm of order $n$
is given by
\be
\label{poly}
{\rm Li}_n(x)=\sum_{k=1}^{\infty} {x^k \over k^n}.
\ee
If we redefine
\be
 h (x)\rightarrow h(x) + {i \theta \over t p} ,
\ee
the functional (\ref{spf}) becomes
\be
\begin{aligned}
S[h]=&-\int_0^1 dx \int_0^1 dy \log | 2 \sinh \, {t\over 2} (h(x)-h(y))|+ {p t \over 2} \int_0^1 dx h(x)^2 \\
&-{pt \over 24} + {\theta^2 \over 2 p t} + {2\over t^2}F_0^{\rm CS}(t).
\end{aligned}
\ee
Since the inclusion of $\theta$ only leads to an additive term in the planar free energy, we will set
$\theta=0$ from now on. After introducing a density function $\rho(h)$ as in (\ref{hdens}),  the $\rho$-dependent part of the effective action can be written (\ref{spf}) as
\be
\label{actionf}
S[\rho]= -\int dh \int dh' \, \rho(h) \rho(h') \log | 2 \sinh \, {t\over 2} (h-h')| + {p t \over 2} \int dh \rho(h)h^2.
\ee
As explained in the previous section, to see if there is a phase transition one first solves for the $\rho(h)$ that extremizes (\ref{actionf}), assuming a one-cut structure. In order to compute $\rho(h)$, we have to solve the integral equation derived from (\ref{actionf}),
\be
\label{inteq}
p h ={\rm P}\,  \int d h' \rho(h') \coth { t\over 2} (h-h'),
\ee
where ${\rm P}$ denotes principal value. The density $\rho(h)$ is supported on a symmetric interval $(-a, a)$.
A similar integral equation appears in the saddle-point analysis of the Chern-Simons matrix model on
the three-sphere \cite{MM,AKMV,MMr}. In fact, after the change of variables $\beta=t h$, (\ref{actionf}) becomes the
planar functional for the Chern-Simons matrix model
\be
\label{csmod}
Z_N=\int\prod_{i=1}^N {d \beta_i \over 2\pi} \,
\prod_{i<j} \Bigl( 2 \sinh {\beta_i - \beta_j\over 2} \Bigr)^2
\exp\Bigl\{-{N \over 2 \xi} \sum_{i=1}^N \beta^2_i \Bigr\},
\ee
with 't Hooft parameter $\xi=t/p$. This connection suggests an effective way of solving
(\ref{inteq}). As in \cite{AKMV,Tierz, MMr}, we change variables
\be
\lambda={\rm e}^{t h +t/p},
\ee
and we introduce the density for the new variable $\lambda$,
\be
\label{transdens}
\rho(\lambda)={d h \over d\lambda}\rho(h)={1\over t \lambda} \rho(h).
\ee
The integral equation (\ref{inteq}) becomes
\be
{1\over 2} {p\over t} {\log \lambda \over \lambda} ={\rm P}\,  \int d \lambda' { \rho(\lambda')
\over \lambda - \lambda'}.
\ee
This is exactly the saddle-point equation for the Chern-Simons/Stieltjes-Wigert matrix model,
and we can solve it in a variety of ways \cite{AKMV,HY,MMr}. The direct computation performed in \cite{MMr} is the most
convenient one in view of the two-cut solution that we will introduce later, so let us briefly review it.
As usual, we introduce a resolvent
\be
\label{resolv}
\omega_0 (\lambda)=\int d \lambda' { \rho(\lambda')
\over \lambda - \lambda'},
\ee
which due to the normalization (\ref{densnorm}) and the redefinition (\ref{transdens}), satisfies the
following asymptotic behaviour
\be
\label{asymres}
\omega_0(\lambda) ={1\over \lambda} + {\cal O}(\lambda^{-2}),
\ee
as $\lambda \rightarrow \infty$. The density $\rho(\lambda)$ is recovered from the resolvent $\omega_0(\lambda)$ through the standard
equation
\be
\label{rhow}
\rho(\lambda) =-{1 \over 2 \pi i} \bigl(\omega_0(\lambda+ i\epsilon) -
\omega_0 (\lambda-i \epsilon)\bigr).
\ee
We are looking for a one-cut solution to the problem, therefore we assume that the density of
eigenvalues is supported in the interval $(a^-,a^+)$, where
\be
a^{\pm} =e^{\pm t a + t/p}.
\ee
It is well known that $\omega_0(\lambda)$ can be computed as \cite{musk}
\be
\label{solwo}
\omega_0(\lambda) =r(\lambda) \oint_{\cal C} {d z \over 2 \pi i} { g(z) \over (\lambda-z) r(z)} ,
\ee
where ${\cal C}$ is a contour around the cut $(a^-,a^+)$, and
\be
g(\lambda)={p \over 2 t } {\log \lambda\over \lambda}, \quad r(\lambda)={\sqrt { (\lambda-a^-)(\lambda-a^+)}}.
\ee
\begin{figure}[!ht]
\leavevmode
\begin{center}
\epsfysize=4cm
\epsfbox{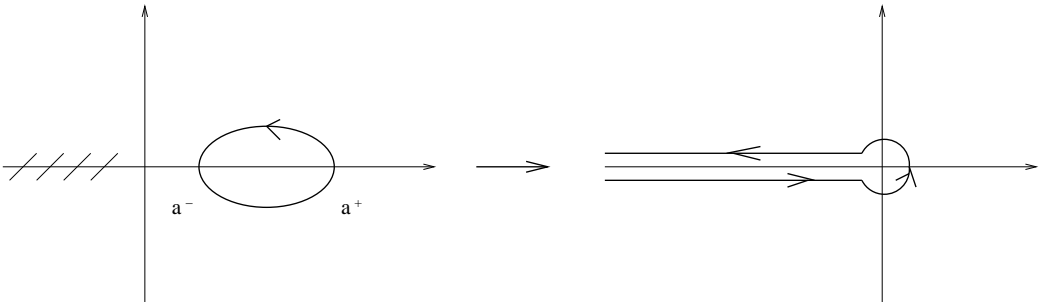}
\end{center}
\caption{This figure shows the deformation of the contour needed to compute the resolvent in (\ref{solwo}). We pick a residue at $z=p$, and
we have to encircle the singularity at the origin as well as the branch cut of the logarithm, which
on the left hand side is represented by the dashed lines.}
\label{contour}
\end{figure}

The standard way to compute an integral like (\ref{solwo}) is to deform the contour. Since the logarithm
has a branch cut, we cannot push the contour to infinity. Instead, we
deform the contour as indicated in \figref{contour}. We pick a pole at $z=\lambda$, and then we
surround the cut of the logarithm along the negative real axis and the singularity
at $z=0$ with a small circle $C_{\epsilon}$ of radius $\epsilon$. A similar situation appears in, for
example, \cite{kazakov}. The final formula for the resolvent is
\begin{equation}
\omega_0(\lambda) = {p\over 2t} {\log \lambda \over \lambda} \\
-{p \over 2t} r(\lambda) \lim_{\epsilon \to 0} \Biggl\{ -\int_{-\infty}^{-\epsilon}{dz \over z(z-\lambda) r(z)}
+ \oint_{C_{\epsilon}} {dz \over 2\pi i}{\log z \over z(z-\lambda) r(z)} \Biggr\}.
\end{equation}
The integrals in the second line have $\log \epsilon$ singularities as $\epsilon \rightarrow 0$, but they
cancel each other, and after
some computations one finds for the resolvent:
\begin{equation}
\begin{aligned}
\omega_0(\lambda)=&-{p \over 2 t \lambda}
\log \Biggl[ { ({\sqrt {a^-}}{\sqrt{ \lambda-a^+}}-{\sqrt {a^+}}{\sqrt {\lambda-a^-}})^2
\over ({\sqrt {\lambda-a^-}}-{\sqrt {\lambda-a^+}})^2 \lambda^2}\Biggr] \\
&+ {p\over 2t \lambda} r(\lambda) {1\over \sqrt{a^-a^+}}
\log \Biggl[ {4 a^-a^+ \over 2 {\sqrt{a^-a^+}} + a^- +a^+} \Biggr].
\end{aligned}
\end{equation}
In order to satisfy the asymptotics (\ref{asymres}) the second term must vanish, and the first one
must go like $1/\lambda$. This implies
\be
\begin{aligned}
4a^-a^+=& 2 {\sqrt{a^-a^+}} + a^- +a^+,\\
 {\sqrt{a^-}}+ {\sqrt{a^+}}=&2 e^{t/p},
 \end{aligned}
\ee
and from here we obtain the positions of the endpoints of the cut $a^-, a^+$ as a function of $t/p$:
\be
\begin{aligned}
\label{endpoints}
a^-=&2e^{2t/p}-e^{t/p} -2e^{3 t \over 2p }{\sqrt {e^{t/p}-1}}, \\
a^+=&2e^{2t/p}-e^{t/p} +2e^{3 t \over 2 p}{\sqrt {e^{t/p}-1}}.
\end{aligned}
\ee

The final expression for the resolvent is then
\begin{equation}
\label{csres}
\omega_0(\lambda)=-{p \over t \lambda}
\log \Biggl[ {1 + e^{-t/p} \lambda+ {\sqrt {(1+e^{-t/p} \lambda)^2-4 \lambda }}  \over 2 \lambda}\Biggr],
\end{equation}
and from here we easily find the density of eigenvalues
\be
\label{swdensity}
\rho(\lambda) =
{p \over \pi t \lambda} \tan^{-1} \Biggl[{ {\sqrt {4 \lambda -(1+e^{-t/p} \lambda)^2 }}  \over
1 + e^{-t/p} \lambda}\Biggr].
\ee
We can now go back to the original variable $h$, to find
\be
\label{density}
\rho(h)={p\over \pi} \tan^{-1} \biggl[ {\sqrt {e^{A/p^2}- \cosh^2 (A h / (2 p))} \over
\cosh(A h /(2p))}\biggr],
\ee
which has its support on $(-a,a)$ with
\be
a={2 p\over A} \cosh^{-1}(e^{A/(2p^2)}).
\ee
As a test of this result, notice that in the double-scaling limit (\ref{dsl}) one finds
\be
\rho(h)= \rho_G(h, 1/A)+ {\cal O}(1/p^2) ,
\ee
therefore the leading term is exactly the Wigner semi-circle distribution obtained by \cite{DK}.

In order to assess the possibility of phase transitions, we have to verify the condition (\ref{rineq}). Notice
first that $|\tan^{-1}(x)|\le {\pi \over 2}$,
therefore
\be
\rho(h)\le p/2
\ee
for all $h$. A first conclusion is that {\it there is no phase transition for} $p\le 2$. For $p>2$ there is indeed a phase transition which occurs when the value of $A$ is such that the maximum of the distribution reaches the value $1$. Since the maximum occurs at $h=0$, we immediately find the following line of critical points:
\be
\label{apcrit}
A_*(p)=p^2 \log \biggl( 1 + \tan^2 \Bigl( {\pi\over p}\Bigr)\biggr), \quad p>2.
\ee
As $p \rightarrow \infty$,
\be
A_*(p) \rightarrow \pi^2,
\ee
in agreement with the result of Douglas and Kazakov (\ref{acrit}). Notice that $A_*(p)$ is a decreasing function of $p$ for $p>2$, and
as $p\rightarrow 2^+$, the critical area increases to infinity. For a given $p$, the small
area phase occurs for $A\le A_*(p)$, and in this phase the
planar free energy is well described by the distribution (\ref{density}).

\begin{figure}[!ht]
\leavevmode
\begin{center}
\epsfysize=6cm
\epsfbox{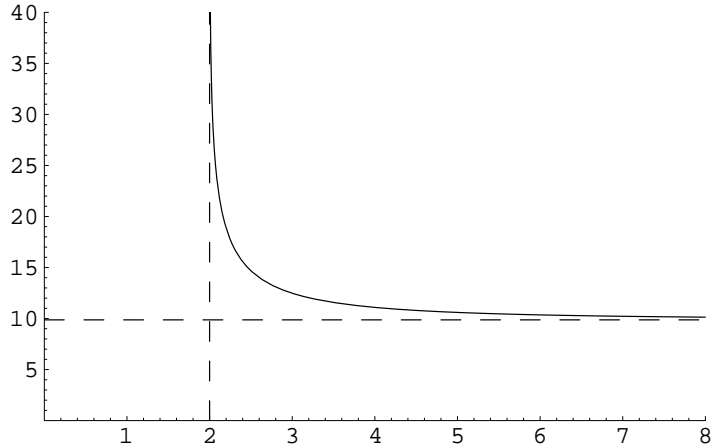}
\end{center}
\caption{This figure represents the phase diagram of $q$-deformed 2d YM theory. The horizontal axis represents the parameter $p$, while the vertical axis represents $A$. 
	The curve shown in the figure is the critical line (\ref{apcrit}),
which separates the phases of small and large area. The horizontal dashed line, which is the asymptote of the curve as $p\rightarrow \infty$,
represents the $A=\pi^2$ critical point of Douglas and Kazakov.}
\label{phased}
\end{figure}
We then have the phase diagram represented in \figref{phased}. The horizontal axis represents the parameter $p$, while the vertical axis represents $A$. The critical line, described by the
function (\ref{apcrit}), has two asymptotes, represented by dashed lines: as $p\rightarrow \infty$ it approaches the horizontal dashed line $A=\pi^2$, which corresponds to the Douglas-Kazakov phase transition. As $p \rightarrow 2^+$ it approaches the vertical asymptote. For $p\in(0,2]$ there is no phase transition. Notice
that, if we parametrize the planar $q$-deformed theory in terms of $p$ and $A$, the region $p\rightarrow \infty$ corresponds to a small deformation, while
the region $p<2$ corresponds to a large deformation. We then see that, if we start with ordinary 2d YM and we turn on the deformation parameter $1/p$,
the Douglas-Kazakov phase transition persists although the critical area increases. At $p=2$ there is a ``barrier" where the critical area becomes infinite. Therefore, when the deformation parameter is large enough, the large $N$ phase transition is smoothed out.

To find the free energy in the small area phase, we
have to compute the functional (\ref{actionf}) evaluated on the density (\ref{density}). Since this functional
is closely related to the functional describing the planar Chern-Simons matrix model, we can borrow
the results from \cite{GV,MMr}. From \cite{MMr} it follows that, at large $N$, the matrix
integral (\ref{csmod}) is given by
\be
\exp\Bigl( N^2 F_0(\xi)\Bigr),
\ee
with
\be
\label{fzero}
F_0(\xi)={1\over \xi^2} F_0^{\rm CS}(\xi)+ {\xi \over 12}.
\ee
Since $\xi=t/p$ in our example, we finally obtain
\be
F^q_0 (t,p)={1\over t^2} \Bigl( p^2 F_0^{\rm CS}(t/p) - 2 F_0^{\rm CS}(t)\Bigr) + {t\over 12 p} + {p t \over 24}.
\ee
As a further check of this expression, notice that, after using the
expansion,
\be
{\rm Li}_3(e^{-t})=\zeta(3) -{\pi^2\over 6} t +\Bigl( {3\over 4} -{1\over 2} \log\, t\Bigr) t^2 + {\cal O}(t^3),
\ee
one finds in the double-scaling limit (\ref{dsl})
\be
F^q_0(t,p) \rightarrow {A\over 24} -{1\over 2} \log\, A +{3 \over 4},
\ee
which indeed is the free energy of the usual 2d YM theory in the small area phase.

\sectiono{Instanton analysis}

Since $q$-deformed 2d YM theory is a one-parameter deformation of the standard one, we expect
the phase transition discovered in the previous section to be triggered by instantons as well. In this section we will verify this by computing the one-instanton suppression factor in the $q$-deformed case. This will also give an intuitive explanation of why the phase transition is absent for $p\le 2$.

The starting point of the discussion is to write the partition function of the theory in a way that makes manifest the instanton content of the model. Since $q$-deformed 2d YM theory has the same
action as standard 2d YM, but differs in the measure \cite{AOSV}, we expect the partition function to be
expressed in terms of a sum over instantons,
\be
\label{semiz}
Z^q=\sum_{n_j}  w(n_j) \exp\biggl( -{2\pi^2 N\over A} \sum_{j=1}^N n_j^2\biggr),
\ee
where $n_j$, $j=1, \cdots, N$, are the instanton numbers characterizing a classical solution \cite{GM},
and $w(n_i)$ is the weight of such a configuration in the semiclassical expansion. In order to compute the weights $w(n_j)$, we follow the technique used by Minahan and Polychronakos \cite{MP} in standard 2d YM and perform a Poisson
resummation of the original expression (\ref{zq}). This can be regarded as a duality transformation which takes us from the large $A$ phase where the expansion (\ref{zq}) is valid, to the small area phase where the semiclassical expansion (\ref{semiz}) is
valid. The partition function can then be written as
\be
Z^q=C\sum_{n_j} F_2 (2 \pi n_j),
\ee
where $F_2(x_j)$ is a Fourier transform with respect to the variables $p_j=l_j-j+1/2$:
\be
F_2(x_j)= \int \prod_j dp_j e^{-i \sum_j x_j p_j} \prod_{j<k} \biggl( 2 \sinh {t \over N}(p_j-p_k)\biggr)^2  \exp\Bigl(-{A\over 2 N} \sum_j p_j^2 \Bigr),
\ee
and we are setting the $\theta$ angle to zero. This transform can be performed by first
computing
\be
\label{fone}
F_1(x_j)= \int \prod_j dp_j e^{-i \sum_j x_j p_j} \prod_{j<k} \biggl( 2 \sinh {t \over N}(p_j-p_k)\biggr)  \exp\Bigl(-{A\over 2 N} \sum_j p_j^2 \Bigr),
\ee
and then doing a convolution. The integral (\ref{fone}) reduces to a Gaussian after using
Weyl's denominator formula for a general Lie algebra,
\be
\sum_{w \in {\cal W}} \epsilon (w) e^{w(\rho)\cdot u} =\prod_{\alpha>0}
2 \sinh {\alpha\cdot u \over 2},
\label{wdf}
\end{equation}
where $\alpha$ are the positive roots, $w\in {\cal W}$ are the elements of the Weyl group, and $\epsilon(w)$ is the
parity of $w$. We find, up to a multiplicative constant,
\be
F_1(x_j)=\exp\Bigl( -{N\over 2 A} \sum_{j=1}^N x_j^2 \Bigr) \prod_{j<k} 2 \sin {t\over 2A} (x_j -x_k),
\ee
and using convolution we finally obtain
\be
\begin{aligned}
F_2(x_j)= &\exp\Bigl( -{N\over 2 A} \sum_{j=1}^N x_j^2 \Bigr)  \\
& \times \int \prod_{j=1}^N dy_j
\prod_{j<k} \biggl( 4 \sin {t\over 2A} (x_{jk} + y_{jk}) \sin {t\over 2A} (x_{jk} - y_{jk}) \biggr)\exp\Bigl( -{N\over 2 A} \sum_{j=1}^N y_j^2 \Bigr) ,
\end{aligned}
\ee
where we introduced the notation $x_{jk}=x_j-x_k$. The instanton weight has then the expression
\be
w(n_j)=\int \prod_{j=1}^N dy_j
\prod_{j<k} \biggl( 4 \sin {t\over 2A} (2\pi n_{jk} + y_{jk}) \sin {t\over 2A} (2 \pi n _{jk} - y_{jk}) \biggr)
\exp\Bigl( -{N\over 2 A} \sum_{j=1}^N y_j^2 \Bigr),
\ee
which is a $q$-deformed version of the result in \cite{MP} for standard 2d YM.

As it was pointed out in \cite{GM}, a precise way to evaluate the importance of instanton contributions to the
partition function is to compare the contribution of the one-instanton term in the semiclassical expansion (\ref{semiz}) to the contribution of the perturbative vacuum. The relative weight of these contributions defines a function $\gamma(A,p)$ as follows
\be
\label{oneisf}
\exp\Bigl[  -{N \over A}  \gamma(A, p)\Bigr]=\exp\biggl( -N{2\pi^2\over A }\biggr){w_1\over w_0} ,
\ee
where the exponent in the right hand side involves the instanton action for $n_1=1, n_{i>1}=0$, and we have denoted
\be
\label{quot}
{w_1\over w_0}={w(1,0,\cdots, 0) \over w(0, \cdots, 0)}.
\ee
We call the function in (\ref{oneisf}) the one-instanton suppression factor. Notice that, as long as $\gamma(A,p)$ is different
from zero, instantons will be suppressed in the large $N$ limit. The suppression is bigger the larger $\gamma(A,p)$ is. In the remaining
of this section, we will compute $\gamma(A,p)$ in the small area phase of $q$-deformed 2d YM, and we will study its properties.

Let us first define the partition function
\be
\label{auxcs}
Z_N=\int \prod_{j=1}^N dy_j
\prod_{j<k} \biggl( 2 \sin {t\over 2A} (y_j-y_k) \biggr)^2
\exp\Bigl( -{N\over 2 A} \sum_{j=1}^N y_j^2 \Bigr).
\ee
This is very close to the partition function of the Chern-Simons matrix model,
although it has a $\sin$ interaction between eigenvalues instead of a $\sinh$ interaction.
We can then use the results of the previous section after changing
\be
\label{imvar}
p\rightarrow -i{p\over A}, \quad A\rightarrow {1\over A},
\ee
and doing carefully the analytic continuation of $p$ to the imaginary axis. Equivalently, we can
change variables $y=-i A \beta/t$ in (\ref{auxcs}) to obtain the matrix model
(\ref{csmod}) with $\xi=-A/p^2$.
One can then see from the formulae presented in the last section that
the planar limit of (\ref{auxcs}) is controlled by the
following density of eigenvalues,
\be
\zeta(y)={p\over \pi A} \tanh^{-1} \biggl[{  {\sqrt { \cos^2(y/(2p))-e^{-A/p^2}}}
\over  \cos(y/(2p))}\biggr],
\ee
with endpoints located at
\be
Y=2p \cos^{-1}(e^{-A/(2p^2)}).
\ee
As $p\rightarrow \infty$, one can easily check that $\zeta(y)\rightarrow \rho_G(y,A)$.

We can now evaluate (\ref{quot}). Notice first that $w_0=Z_N$. On the other
hand, as in \cite{GM}, one has
\be
\label{wonex}
\begin{aligned}
&w_1=\\
&Z_{N-1} \int_{-\infty}^{\infty} dy_1 e^{-{N \over 2 A} y_1^2} \Big\langle \prod_{j=2}^N \biggl( 4 \sin {1\over 2p}\bigl(2\pi + (y_j-y_1)\bigr) \sin {1\over 2p}\bigl(2\pi - (y_j-y_1)\bigr)\biggr) \Big\rangle_{N-1},
\end{aligned}
\ee
where the correlator is computed in the model (\ref{auxcs}) with $N-1$ variables.
Since we are interested in the large $N$ behavior of the one-instanton suppression
factor, we can compute the different integrals in the saddle-point approximation. This in particular means that we can set $y_1=0$ inside
the correlator in (\ref{wonex}). We find,
\be
\begin{aligned}
&{w_1\over w_0}=\\
&\biggl( {2 \pi A \over N} \biggr)^{1/2} {Z_{N-1} \over Z_N}
\exp \biggl\{ (N-1) \int dy \zeta(y) \log \Bigl( 4 \sin {1\over 2p}(2\pi + y) \sin {1\over 2p}(2\pi - y) \Bigr) \biggr\}.
\end{aligned}
\ee
We have first to evaluate the quotient $Z_{N-1}/Z_N$ in the large $N$ limit. It is easy to see that, at leading
order in $N$, this quotient is
\be
\exp \Bigl\{-N ( 2 F_0(\xi) + \xi F_0'(\xi) )\Bigr\},
\ee
where $F_0(\xi)$ is given in (\ref{fzero}). Here, $\xi=-A/p^2$, and after an analytic continuation
$\xi \rightarrow -\xi$ we find,
\be
2 F_0(\xi) + \xi F_0'(\xi)= {p^2\over A}\Bigl( {\rm Li}_2(e^{-A/p^2})-{\pi^2\over 6}\Bigr),
\ee
up to an overall sign $(-1)^N$ in $Z_{N-1}/Z_N$. Putting everything together, we obtain the following formula for the
function $\gamma(A,p)$ defined in (\ref{oneisf}):
\be
\label{is}
\begin{aligned}
\gamma(A,p)=&2 \pi^2  + p^2 \Bigl( {\rm Li}_2(e^{-A/p^2})-{\pi^2\over 6} \Bigr) \\
&- A \int dy \zeta(y) \log \Bigl( 4 \sin {1\over 2p}(2\pi + y) \sin {1\over 2p}(2\pi - y) \Bigr).
\end{aligned}
\ee
The integral in (\ref{is}) can be evaluated analytically. Notice first that in any matrix model one has
\be
\label{reslog}
F(v)\equiv \int d\lambda \rho(\lambda) \log\bigl(1-{\lambda/v}\bigr)=\int_{\infty}^v dv' \bigl( \omega_0(v') -1/v'\bigr).
\ee
This follows directly from the definition of the resolvent in (\ref{resolv}). Taking into account the redefinition (\ref{imvar}), we find
that the integral in (\ref{is}) is given by
\be
2 {\rm Re} \, F(e^{- A/p^2 + 2\pi i/p}),
\ee
where $F(v)$ is obtained as in (\ref{reslog}), and the relevant resolvent is (\ref{csres}). After some work, and
using standard identities for the dilogarithm, one finds the following expression:
\be
\begin{aligned}
\label{finalg}
\gamma(A, p)=&2 \pi^2 -p^2 \Bigl( {\rm Li}_2(e^{-A/p^2})+{\pi^2\over 6} \Bigr)+
2 p^2  {\rm Re}\, {\cal G}(f_+(p,A), f_-(p,A)),
\end{aligned}
\ee
where
\be
\begin{aligned}
{\cal G}(x,y)=&{1\over 2} (\log \, x)^2 + \log\, x \log\, (1-y) + {\rm Li}_2(1-x) + {\rm Li}_2(y),\\
f_{\pm}(p,A)=&\exp\Bigl(\pm A/(2 p^2) + i(\varphi - \pi/p)\Bigr),\\
\varphi=& \tan^{-1}\biggl( {\sqrt {e^{-A/p^2} -\cos^2(\pi/p)}\over \cos(\pi/p)}\biggr).
\end{aligned}
\ee
\begin{figure}[!ht]
\leavevmode
\begin{center}
\epsfysize=6cm
\epsfbox{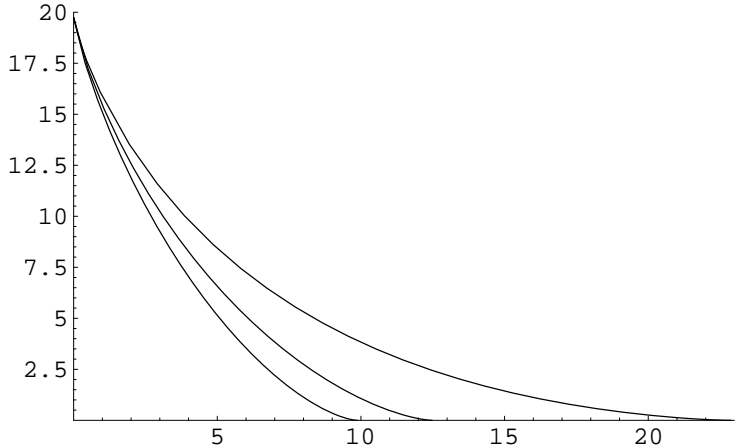}
\end{center}
\caption{This figure shows the function $\gamma(A,p)$ appearing in the one-instanton suppression factor, plotted as a function of $A$, and
for the values $p=2.1, 3, \infty$, from top to bottom.
For each $p$ it is a decreasing function of the area and vanishes at the critical value $A_*(p)$.}
\label{insts}
\end{figure}
In order to understand the properties of the instanton suppression factor, we have studied (analytically and numerically) the properties of
(\ref{finalg}) as a function of $A$ and $p$ for $p>2$, $A\le A_*(p)$. The main results of this analysis are the following:

1. As $p\rightarrow \infty$, the function $\gamma (A, p)$ becomes the function $\gamma_{\rm GM}(A)$
introduced in (\ref{grossm}). This is a consistency check of the solution.

2. For any fixed $p>2$, $\gamma(A,p)$ takes the value $2 \pi^2$ at $A=0$ and then it decreases monotonically as the area is increased. At the critical area (\ref{apcrit}) one has
\be
\label{vanish}
\gamma (A_*(p),p)=0.
\ee
The vanishing of $\gamma(A, p)$ at the critical area can be proved analytically, since at $A=A_*(p)$,
\be
f_{\pm}(p,A_*(p))=\Bigl( \cos (\pi/p)\Bigr)^{\mp 1} e^{-i\pi/p}.
\ee
For arguments of this form (which are algebraic numbers)
the dilogarithm satisfies nontrivial identities \cite{K} that can be easily shown to lead to (\ref{vanish}).

3. For $p<p'$, one has that $\gamma(A,p) >\gamma(A,p')$ in their common range $A\le A_*(p')$.

These properties are illustrated in \figref{insts}, which shows the function $\gamma(A,p)$ as a function of the area for the values $p=2.1, 3, \infty$, from top to bottom.
The above properties show that the one-instanton suppression factor in the small area phase decreases as the area grows, until it vanishes at $A_*(p)$. Therefore, at the line of critical points found in section 3, the instantons are not suppressed anymore and they become favorable configurations. This shows that the phase transition for the $q$-deformed theory is indeed triggered by instantons, and follows a mechanism similar
to the one studied in \cite{GM}: for $A> A_*(p)$, the entropy of the instantons dominates over their Boltzmann weight. The above analysis also
shows that, as $p$ decreases, the instanton suppression factor becomes bigger and bigger, pushing the critical value of
the area to ever larger values. This indicates that the smoothing out of the phase transition for $p \le 2$ is due to the fact that the
instantons are suppressed for all values of $A$ and we only have one phase dominated by the perturbative vacuum $n_i=0$.

\sectiono{The two-cut solution}

In this section we give some preliminary results about the large area phase of the theory. After the phase transition found in section 3,
we expect a distribution $\rho(h)$ a la Douglas-Kazakov, with the shape shown in the r.h.s. of \figref{dkt} and characterized by two points $\ha, \hb$. The distribution governing the large area distribution is then of the form
\be
\label{rhom}
\rho(h)=\begin{cases} \tilde \rho(h), &\text{$-\ha \le h \le -\hb$,  $\hb \le h\le \ha$}, \\
1, &\text{$-\hb \le h \le \hb$.}
\end{cases}
\ee
After changing variables $\lambda=\exp(th + t/p)$ as in the previous section, the new density of eigenvalues $\tilde \rho (\lambda)=1/(t \lambda) \tilde \rho (h)$ has support on the two intervals
$(a^-, b^-)$, $(b^+, a^+)$, where
\be
a^{\pm}=e^{t/p \pm t\ha}, \quad  b^{\pm}=e^{t/p \pm t\hb}.
\ee
This density satisfies
the following integral equation,
\be
g(\lambda)\equiv \frac{p}{2t}\; {\log \lambda \over \lambda}
+{1\over t  \lambda} \log { \lambda/b^+ -1\over \lambda/b^- -1}
 ={\rm P} \int \frac{\tilde{ \rho}(\lambda^{\prime} )}{\lambda - \lambda^{\prime} } d \lambda^{\prime}
\ee

\begin{figure}[!ht]
\leavevmode
\begin{center}
\epsfysize=4cm
\epsfbox{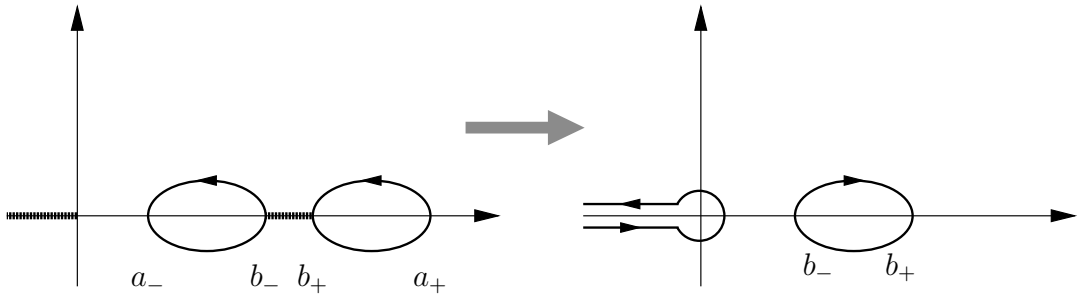}
\end{center}
\caption{This figure shows the deformation of the contour needed to compute the resolvent in
the two-cut solution. We have to encircle the singularity at the origin, and the two branch cuts
denoted by thick lines on the left.}
\label{tcontour}
\end{figure}
As in the one-cut case, we introduce a resolvent
\be
\tilde \omega_0(\lambda)=\int {\tilde \rho(\lambda') \over \lambda -\lambda'} d \lambda^{\prime}.
\ee
This can be again computed by the contour integral (\ref{solwo}), but now
\be
r(z)={\sqrt {(z-a^-)(z-a^+)(z-b^-)(z-b^+)}}.
\ee
and ${\cal C}$ is the union of the contours surrounding the cuts $(a^-, b^-)$, $(a^+, b^+)$. To perform
the integral (\ref{solwo}) we deform the contours in the way shown in \figref{tcontour}: we now encircle the branch cut along $(-\infty, 0)$, coming from $\log \lambda$, and the branch cut of the integrand along $(b^-, b^+)$. The answer
for the resolvent is
\be
\begin{aligned}
\label{tcresolvent}
\tilde \omega_0(\lambda)=&{p\over 2 t} {\log \lambda \over \lambda} +{1 \over t \lambda} \log { \lambda/b^+-1
\over \lambda/b^--1} \\
& -r(\lambda) {p\over 2 t} \lim_{\epsilon \to 0} \biggl\{ -\int_{-\infty}^{-\epsilon} {dz \over z (z-\lambda) r(z) } + \oint_{C_{\epsilon}}
{dz \over 2 \pi i} { \log z \over z (z-\lambda) r(z) } \biggr\}
\\
&+  {r(\lambda)\over t} \int_{b^-}^{b^+} {dz \over z (z-\lambda) r(z) }.
\end{aligned}
\ee
The above integrals can be expressed in terms of elliptic functions. We will
change notation $a^-,b^-, b^+, a^+$ to $d,c,b,a$. Define
\be
\begin{aligned}
I(\lambda, u)&\equiv \int_u^d {dz \over (z-\lambda) r(z)} \\
&={2 \over (\lambda-c)(\lambda-d) {\sqrt {(a-c)(b-d)}}}
\biggl\{ (c-d) \Pi (\phi, n, k) + (d-\lambda) F(\phi, k) \biggr\},
\end{aligned}
\ee
where $\Pi(\phi,n,k)$ and $F(\phi,k)$ are incomplete elliptic integrals of the third and the first kind,
respectively, and
\be
\sin^2 \phi= {(a-c)(d-u) \over (a-d)(c-u)}, \quad n={(a-d)(\lambda-c) \over (a-c) (\lambda -d)}, \quad
k^2 = {(b-c)(a-d)\over (a-c)(b-d)}.
\ee
In what follows it will be convenient to introduce the following angles $\phi_1$, $\phi_2$
and variables $n_1$ and $n_2$:
\be
\label{nangles}
\begin{aligned}
\sin^2 \phi_1= &{a-c\over a-d}, \quad \sin^2 \phi_2={d\over c}
 {a-c\over a-d}, \\
 n_1=&{a-d \over a-c}, \quad n_2={c\over d} {a-d\over a-c}.
 \end{aligned}
 \ee
 In terms of these variables one finds,
 \be
 \begin{aligned}
 I(\lambda, -\infty)=&{2 \over (\lambda-c)(\lambda-d) {\sqrt {(a-c)(b-d)}}}
\biggl\{ (c-d) \Pi (\phi_1, n, k) + (d-\lambda) F(\phi_1, k) \biggr\}, \\
 I(\lambda, 0)=&{2 \over (\lambda-c)(\lambda-d) {\sqrt {(a-c)(b-d)}}}
\biggl\{ (c-d) \Pi (\phi_2, n, k) + (d-\lambda) F(\phi_2, k) \biggr\}, \\
 I(0, -\infty)=&{2 \over c d {\sqrt {(a-c)(b-d)}}}
\biggl\{ (c-d) \Pi (\phi_1, n_2, k) + d F(\phi_1, k) \biggr\}.
\end{aligned}
\ee
The first integral in the second line of (\ref{tcresolvent}) is given by
\be
{1\over \lambda} \Bigl( I(\lambda, -\infty) -I(\lambda,0) - I(0, -\infty)+I(0, -\epsilon)\Bigr).
\ee
The second integral in the second line is simply a residue and it can be computed immediately:
\be
\label{logsing}
-{1\over \lambda} {1\over {\sqrt {a b c d}}} \log \epsilon.
\ee
We now compute $ I(0, -\epsilon)$ at next-to-leading order in $\epsilon$. This will have
a logarithmic singularity
which will cancel (\ref{logsing}). In order to do that, we need the following identity \cite{bateman}:
\be
\label{tkiden}
\Pi(\phi,n, k)=\delta(n) \biggl\{ {1\over 2}\log
{\vartheta_1(v + \beta) \over \vartheta_1 (v-\beta)} -{\vartheta_4'(\beta) \over
\vartheta_4(\beta)} v\biggr\},
\ee
where
\be
\delta(n)=\biggl( {n \over (1-n)(k^2 -n)} \biggr)^{1\over 2}, \quad
v={F(\phi,k) \over 2 K(k)}, \quad \beta={F(\sin^{-1}(n^{-{1\over 2}}), k)\over 2 K(k)},
\ee
and the $\tau$ parameter in the theta functions is given as usual by
\be
q=e^{2\pi i \tau}=\exp(-\pi K'(k)/K(k)).
\ee
Notice that, when
\be
\sin^2\, \phi ={1\over n}
\ee
we have a logarithmic singularity in the elliptic integral $\Pi(\phi,n,k)$. This is immediately checked
in the integral representation of the elliptic function\footnote{Notice that in the conventions we are using
the $n$ in $\Pi(\phi,n,k)$ corresponds to $-n$ in the definition given in \cite{bateman}.}. We can now use (\ref{tkiden}) to extract the next-to-leading
behavior. Since
\be
\sin^2 \phi={a-c\over a-d} {d+\epsilon \over c+\epsilon}, \quad n= n_2,
\ee
the leading behaviour of $\Pi(\phi,k,n_2)$ is given by
\be
\label{ellipticPi}
\begin{aligned}
& \delta(n_2)\biggl( -\beta_2 {\vartheta_4'(\beta_2) \over
\vartheta_4(\beta_2)} + {1\over 2} \log {\vartheta_1(2\beta_2) \over \vartheta_1'(0)}  -
{1\over 2} \log \Bigl( { c-d \over 4 c d} \delta (n_2)\Bigr)  +
{1\over 2} \log \, K(k) \biggr) \\
&-{\delta(n_2)\over 2} \log \epsilon + {\cal O}(\epsilon),
\end{aligned}
\ee
where $\beta_2$ is given by
\be
\beta_2={F(\phi_2, k)\over 2 K(k)}.
\ee
This leads to the following expression
 \be
 I(0, -\epsilon) =-{1\over  {\sqrt {a b c d}}} \log \epsilon + I(0,0) +{\cal O}(\epsilon),
 \ee
 where
 \be
 \begin{aligned}
 I(0,0)=&{1 \over {\sqrt {a b c d}}}\biggl( -2 \beta_2 {\vartheta_4'(\beta_2) \over
\vartheta_4(\beta_2)} + \log {\vartheta_1(2\beta_2) \over \vartheta_1'(0)}  +
 \log \Bigl( { c-d \over 4 c d} \delta (n_2)\Bigr)  -
 \log \, K(k) \biggr) \\
&+ {2 \over c {\sqrt {(a-c)(b-d)}}} F(\phi_2, k).
\end{aligned}
\ee
From the above result we see that the singularities as $\epsilon \rightarrow 0$ cancel, as wished.

We now consider the remaining integral. Define
\be
\begin{aligned}
J(\lambda)\equiv& \int_c^b {dz \over (z-\lambda) r(z)} \\
&={2 \over (\lambda-a)(\lambda-b) {\sqrt {(a-c)(b-d)}}}
\biggl\{ (a-b) \Pi ( m, k) + (b-\lambda) K( k) \biggr\},
\end{aligned}
\ee
where
\be
m={(b-c)(\lambda-a) \over (a-c) (\lambda -b)}.
\ee
We then have,
\be
\int_{c}^{b} {dz \over (z-\lambda) z r(z) }={1\over \lambda}\Bigl( J(\lambda)-J(0)\Bigr),
\ee
where $J(0)$ is given explicitly as
\be
J(0)
={2 \over a b  {\sqrt {(a-c)(b-d)}}}
\biggl\{ (a-b) \Pi ( m(0), k) + b  K( k) \biggr\}.
\ee
Putting everything together, we find the following expression for the resolvent:
\be
\label{tcresint}
\begin{aligned}
\tilde \omega_0 (\lambda)=&{p\over 2 t} {\log \lambda \over \lambda} +{1 \over t \lambda} \log { \lambda/b-1
\over \lambda/c-1} \\
& + {p r(\lambda)\over 2 t \lambda} \Bigl( I(\lambda, -\infty) -I(\lambda,0) - I(0, -\infty)+ I(0, 0)\Bigr)\\
&+  {r(\lambda)\over  t \lambda}\Bigl(J(\lambda)-J(0)\Bigr).
\end{aligned}
\ee
As $\lambda \rightarrow \infty$, this is indeed a Laurent series in $\lambda$: using again (\ref{tkiden}) it is easy to see that $I(\lambda, -\infty)$
contains a term of the form $-\log(\lambda)/r(\lambda)$ that cancels against the first term in (\ref{tcresint}). In order to derive the conditions for the endpoints of the cut, we must impose the asymptotic behaviour
\be
\label{modas}
\tilde \omega_0(\lambda)= {1-2\hat b \over \lambda} +{\cal O}(\lambda^{-2}).
\ee
We find three conditions. First of all, notice that there is a term of order $\lambda$ coming from the integrals $I(0,0)$, $I(0, -\infty)$, and
$J(0)$. Imposing the cancellation of this term, one obtains the condition
\be
\label{first}
p (I(0,0) -I( 0, -\infty)) -2 J(0)=0.
\ee
The vanishing of the constant term leads to the condition
\be
\label{second}
p\Bigl( F(\phi_2, k) -F(\phi_1, k)\Bigr) = 2K(k),
\ee
Finally, the fact that the $1/\lambda$ term has the coefficient $1- 2\hb$ leads to a third condition,
\be
\label{third}
\begin{aligned}
& p\biggl( (a+b+d-c)(F(\phi_1, k) -F(\phi_2, k)) -2 (c-d)  \Pi (\phi_2, n_1, k) \\
&+ {\sqrt {(a-c)(b-d)}} \Bigl(- 2 \beta_1 {\vartheta_4'(\beta_1) \over
\vartheta_4(\beta_1)} + \log {\vartheta_1(2\beta_1) \over \vartheta_1'(0)}  -
 \log \Bigl( { c-d \over 4 } \delta (n_1)\Bigr)  -
 \log \, K(k) \Bigr) \biggr) \\
& +2  (b+d+c-a) K(k) + 2 (d-b) \Pi (m_{\infty}, k) =t,
\end{aligned}
\ee
where
\be
\beta_2={F(\phi_2, k)\over 2 K(k)}, \qquad m_{\infty}={ b-c\over a-c}.
\ee
These conditions determine the endpoints $\hat a, \hat b$ as functions of the parameters $t,p$. We seem to have three conditions for
two unknowns, but since we started with a symmetric problem and we just changed variables, one of the conditions is redundant. This is not
easy to verify from the above expressions, but can be checked, for example, by doing a small $t$ expansion of the equations, and
assuming a power series ansatz for the endpoints:
\be
\hat a(t,A)=\sum_{n=0}^{\infty} \ha_n(A) \, t^n, \quad  \hat b(t,A)=\sum_{n=0}^{\infty} \hb_n(A) \, t^n.
\ee
The ansatz is justified by the fact that, as $t\rightarrow 0$ with $A$ fixed, we must recover the standard YM result obtained in \cite{DK}.
One can see that, at leading order in $t$, the three conditions above lead to the same equation, namely
\be
{\ha_0 + \hb_0 \over 2}A =2 K(k_0),
\ee
where
\be
k_0^2={4 \ha_0 \hb_0 \over (\ha_0 + \hb_0)^2}.
\ee
Using standard properties of elliptic functions, one can easily check that this condition becomes
\be
\label{firstDK}
A={ 4 \over \ha_0} K(\hb_0/\ha_0),
\ee
which is precisely one of the equations found in \cite{DK}.  Notice that, by making use of (\ref{first}), we can simplify the expression of the resolvent
to
 \be
\label{tcresfinal}
\begin{aligned}
\tilde \omega_0 (\lambda)=&{p\over 2 t} {\log \lambda \over \lambda} +{1 \over t \lambda} \log { \lambda/b-1
\over \lambda/c-1} \\
& + {p r(\lambda)\over 2 t \lambda} \Bigl( I(\lambda, -\infty) -I(\lambda,0) \Bigr)+ {r(\lambda)\over  t \lambda}J(\lambda).
\end{aligned}
\ee
In principle, the above conditions for $\ha, \hb$, together with the explicit expression for the resolvent in (\ref{tcresfinal}), determine completely the
solution for the large area phase. These conditions are rather intricate to be treated analytically, but one could study them numerically.

The most important question to address is the order of the phase transition for different values of $p$. This of course can be seen, as in \cite{DK}, by computing the free energy in the large area phase that we have just analyzed. Since the line of critical points is smoothly connected to the Douglas-Kazakov transition, we should expect the transition in the $q$-deformed theory to be of third order for any $p>2$. Indeed, one can find
indirect evidence that this is the case by using an argument in \cite{GM} based on double-scaling limits. If we consider a theory with a large $N$ $n$-th order phase transition at a critical area $A=A_*$ between phases ${\rm I}$ and ${\rm II}$, the free energy has the following behaviour
\be
F^{\rm I}_0(A)-F_0^{\rm II}(A)\sim (A_*-A)^n.
\ee
To define a double-scaling limit of such a theory, one should introduce a string coupling constant $\mu_s$ through
\be
\mu_s^{-2} =N^2(A_*-A)^n.
\ee
The nonperturbative effects of such a theory are expected to be of the form $\exp(-1/ \mu_s)$. But this means that the instanton effects in the original theory should have the behaviour $\exp(-N\gamma(A))$, with
\be
\label{critg}
\gamma(A)\sim(A_*-A)^{n/2}.
\ee
Indeed, in \cite{GM} it is found that the function (\ref{grossm}) appearing in the instanton suppression factor has exactly the
behaviour (\ref{critg}) with $n=3$ near the Douglas-Kazakov transition point, as required for the existence of
a double-scaling limit at a third order phase transition. According to this argument, the behaviour of the instanton
suppression factor near the critical point can be indeed regarded as an indirect way to probe the order of the phase transition. We have checked numerically that the function $\gamma(A,p)$ that we found in (\ref{finalg}) behaves indeed as
\be
\gamma(A,p)\sim (A_*(p) -A)^{3/2}
\ee
near $A_*(p)$, for various values of $p>2$. This is indeed consistent with the large $N$ phase transition of the $q$-deformed
theory being of third order for all $p>2$.

We should also mention that, in \cite{irantwo}, general criteria have been formulated to determine the order of a phase
transition for a model based on a distribution of Young tableaux. These criteria only depend on the behaviour of the density (\ref{density}) in the
small area phase. It can be easily seen that according to these criteria, the phase transition of the
$q$-deformed theory is of third order for any $p>2$.

\sectiono{Implications for topological string theory and open problems}

In this paper we have shown that $q$-deformed 2d YM theory exhibits an interesting phase structure, with a Douglas-Kazakov phase transition
smoothly connected to that of the standard YM theory, and a ``barrier" at $p=2$. One of the original motivations of this analysis was the appearance of the $q$-deformed theory as a nonperturbative completion of topological string theory on certain Calabi-Yau backgrounds. $q$-deformed
2d YM on the sphere has been proposed in \cite{V, AOSV} as a nonperturbative, holographic description of topological strings on the local Calabi-Yau manifold
\be
\label{localcy}
{\cal O}(-p) \oplus {\cal O}(p-2) \rightarrow {\bf P}^1,
\ee
where the integer number $p>0$ corresponds to the parameter $p$ appearing in (\ref{zq}). Explicit computations in \cite{AOSV} show that the perturbative partition function computed in \cite{BP} appears as a certain decoupling limit of the large area expansion of the $q$-deformed theory.
However, the fact that this theory exhibits a phase transition suggests that, for geometries of the form (\ref{localcy}) with $p>2$, the large area
expansion has a finite radius of convergence which, in terms of the `t Hooft parameter $t$, is given by $t_*(p) =A_*(p)/p$. As $p$ becomes larger,
the radius of convergence becomes smaller. Therefore, the conjecture of \cite{OSV} suggests that for the geometries (\ref{localcy}) with
$p>2$, there will be a phase transition at small radius in the full, nonperturbatively completed topological string theory.
What are the possible interpretations of this phase transition in the topological string theory context? We will mention here three
possibilities, although a better understanding of the implications
of the phase transition of $q$-deformed YM to nonperturbative topological strings will require a more detailed treatment\footnote{The discussion that follows benefited greatly from conversations with M. Aganagic, D. Morrison, H. Ooguri and N. Saulina.}:

1. A first possibility is that the phase transition in the $q$-deformed theory indicates
a topology change in the Calabi-Yau background. After all, the small and the large area phases are described by different master fields of the two-dimensional theory, corresponding to the one-cut and two-cut solutions discussed above, and it is known that in large $N$ dualities the master field encodes
the geometry of the target \cite{DV,AKMV}. This topology change might be also interpreted, as in \cite{DGOV}, in terms of a process involving a
splitting of baby universes.

2. A second possibility is that the small area phase does not have a geometric interpretation. One indication of that
is the string description of standard 2d YM: the analysis of \cite{G,GT} shows that the large area expansion
has an interpretation in terms of branched coverings of the sphere. However, it has been argued that the existence of a large $N$ phase transition suggests
that this geometric picture does not hold for the small area phase \cite{GM}. In the same vein, it is likely that the small area phase of the
$q$-deformed theory is not described appropriately by topological strings with a geometric target. This is in fact very reminiscent of the analysis of
\cite{ag} (see also \cite{bw,agb}), where it was shown that the large $N$ phase transition of the unitary matrix model corresponds, in AdS/CFT at finite
temperature, to the point where the horizon of the
small AdS black hole becomes comparable to the string scale. At this point, the supergravity/geometry picture breaks down. The situation we are
considering here could be a topological string analogue of the large $N$ transition of \cite{ag}.

3. A more conservative possibility is that the conjecture of \cite{OSV} does not fully apply to the local geometries (\ref{localcy})
when $p>2$, or at least does not apply to the small area phase. The original conjecture was formulated for compact Calabi-Yau threefolds, and there may be subtleties when applying it to the noncompact case. It turns out that precisely for $p>2$
there are obstructions for contracting the ${\bf P}^1$ inside (\ref{localcy}) to a point \cite{L, KM}, and because of this reason one can expect
these geometries not to arise as a decompactification limit of a compact Calabi-Yau. It is intriguing that the
``barrier" $p=2$ that we found in this paper is the same that occurs in the geometric setting.

In extracting the consequences of our analysis for the nonperturbative physics of topological strings, there is
another point that should be mentioned. In our analysis we considered the saddle-point solution of the functional $S[h]$, and we found that this leads
to a distribution where $\langle h\rangle=0$ and the dependence on the $\theta$ angle is trivial. However, it has been
argued in \cite{MY}, by studying the instanton weight factors, that the presence of a nonzero $\theta$ changes the
location of the critical line. This is an interesting possibility and deserves further study.
Also, we have restricted ourselves to solutions with zero $U(1)$ charge. This is indeed
the true vacuum of the theory \cite{iran}, but one could
also consider saddle-point solutions like those in \cite{MP}: one imposes the
constraint $\langle h\rangle=Q$, where $Q$ is the $U(1)$ charge, solves for the density, and then
finally sums over all integer charges with a weight $\exp(iQ\theta)$. It may happen that, in order to compare
our results with
those of \cite{AOSV}, one should use this prescription to include the $U(1)$ charges.

It is also worth pointing out
that the instanton weight factors considered in section 4 are closely related to the degeneracies of BPS states analyzed in \cite{AOSV}. It is likely
that the techniques of \cite{GM} that we used and extended to the $q$-deformed case in order to compute these weights lead to a useful
technique to obtain the degeneracies.

From the point of view of the two-dimensional gauge theory, the results of this paper indicate that, when the deformation parameter is
sufficiently large, the large $N$
phase transition is smoothed out already at the planar level. This is an interesting, new mechanism for smoothing out large $N$ transitions which may have implications in other contexts (the other mechanism we are aware of to smooth out these transitions requires performing a double-scaling limit, as in \cite{liu,ag}, and involves a resummation of the $1/N$ expansion).

There are also various open questions concerning the gauge theory aspects of our analysis. Of course, the two-cut solution that we presented in this paper should be investigated in more detail. One could also investigate the phase structure and free energy of the chiral version of the $q$-deformed theory (in the 2d YM case, this has been done in \cite{CT,kostov}). Since the chiral sector makes a more direct contact with the perturbative topological string
amplitudes, this may help in
understanding better the holographic description proposed in \cite{V,AOSV}.  It would be also very interesting to analyze the subleading $1/N$ corrections to the planar result in the small area phase. In \cite{GM} this was done for the standard YM case by using a discretized version of orthogonal polynomials, but
it is not obvious how to generalize this to a discrete model with a sinh interaction. Such a generalization
would make it also possible to define a double-scaled theory near
the critical line of the $q$-deformed theory, as we briefly discussed in the last section.

\section*{Acknowledgments}
We would like to thank Mina Aganagic, David Morrison, Andy Neitzke, Hirosi Ooguri, Natalia Saulina and Erik Verlinde for helpful discussions.
M.M. would also like to thank Luis \'Alvarez-Gaum\'e and Spenta Wadia for conversations on related topics during the last year. M.M. and A.S. would like to
thank the KITP at Santa Barbara, as well as the organizers of the program {\it Mathematical structure of string theory}, for their hospitality and the
great research environment during the completion of this work. While at KITP, their research was supported by the NSF under
grant PHY99-07949. The work of X.A., R.B. and A.S. has been partially supported by Stichting FOM.

\end{document}